\documentclass[a4paper,11pt]{article}
\usepackage{pos}

\title{Disconnected contributions to the magnetic polarisability of the neutral pion}

\author*[a,b]{Ryan Bignell}
\author[b]{Waseem Kamleh}
\author[b]{Derek Leinweber}

\affiliation[a]{Department of Physics, College of Science, Swansea University,\\
  Swansea SA2 8PP, United Kingdom
}

\affiliation[b]{Special Research Centre for the Subatomic Structure of Matter (CSSM),\\
Department of Physics, University of Adelaide, Adelaide, South Australia 5005, Australia
}

\emailAdd{ryan.bignell@swansea.ac.uk}
\emailAdd{waseem.kamleh@adelaide.edu.au}
\emailAdd{derek.leinweber@adelaide.edu.au}
\abstract{
The magnetic polarisability of the neutral pion has been calculated using the background field method in lattice QCD. These early results do not consider the effect of the disconnected loop contractions arising from the breaking of charge symmetry in a background magnetic field. Recent work in chiral perturbation theory has shown that these quark-self-annihilation contractions provide the leading loop-order contributions to the magnetic polarisability. A first investigation of these contractions in a background magnetic field is presented.
}

\FullConference{%
 The 38th International Symposium on Lattice Field Theory, LATTICE2021
  26th-30th July, 2021
  Zoom/Gather@Massachusetts Institute of Technology
}

\usepackage{braket}            
\usepackage{graphicx}          
\usepackage[utf8]{inputenc}    
\usepackage{hyperref}          
\usepackage{dcolumn}           
\usepackage{color}             
\usepackage{adjustbox}         
\usepackage{subcaption}        
\usepackage{csquotes}        

\newcommand{\mbZ}{\mathbb{Z}}
\newcommand{\etal}{\textit{et al.}}

\newcommand{\qud}{q_{u|d}}
\newcommand{\qudb}{\overline{q}_{u|d}}
\newcommand{\gf}{\gamma_{5}}

\newcommand{\abs}[1]{\left|#1\right|}
\newcommand{\order}[1]{\mathcal{O}\left(#1\right)}
\newcommand{\aqeb}{\abs{qe\,B}}

\newcommand{\rb}[1]{\left(#1\right)}
\newcommand{\mpi}{m_\pi}
\newcommand{\qu}{q_{u}}
\newcommand{\qub}{\overline{q}_u}
\newcommand{\qd}{q_{d}}
\newcommand{\qdb}{\overline{q}_d}
\newcommand{\pip}{{\pi^{+}}}
\newcommand{\piz}{{\pi_{0}}}

\newcommand{\eqnrtwo}[2]{Eqs.~$\left(\ref{#1}\right)$ and $\left(\ref{#2}\right)$}
\newcommand{\Fig}[1]{Figure~\ref{#1}}

\newcommand{\Sec}[1]{Section~\ref{#1}}

\newcommand{\Refl}[1]{Ref.~\cite{#1}}    
\newcommand{\Refltwo}[2]{Refs.~\cite{#1} and \cite{#2}}    
\newcommand{\eqnr}[1]{Eq.~$\left(\ref{#1}\right)$}

\begin{document}
\maketitle

\section{Introduction}
The response of the structure of the pion to an external electric or magnetic field is characterised by the electric $\rb{\alpha_\pi}$ and magnetic $\rb{\beta_\pi}$ polarisabilities. These are experimentally measured using Compton scattering experiments~\cite{Antipov:1982kz,Adolph:2014kgj,Filkov:2018cey,Moinester:2019sew} where the polarisabilities describe the scattering angular distribution~\cite{Moinester:2019sew,Holstein:1990qy,Moinester:1997um,Scherer:1999yw,Ahrens:2004mg}.
\par
Pion polarisabilities have been studied in a variety of theoretical frameworks; including chiral perturbation theory~\cite{Gasser:2006qa,Ahrens:2004mg,Burgi:1996qi}, the linear $\sigma$ model~\cite{Bernard:1988gp} and dispersion sum rules~\cite{Filkov:1998rwz,Filkov:2008uyj,GarciaMartin:2010cw}. Close agreement with experiment is possible using chiral perturbation theory~\cite{Gasser:2006qa}. The \emph{ab initio} method for low energy QCD known as lattice QCD has also been to used to calculate electric and magnetic polarisabilities of the pion~\cite{Luschevskaya:2015cko,Bali:2017ian,Bignell:2019vpy,Bignell:2020dze,Ding:2020hxw,Niyazi:2021jrz,Wilcox:2021rtt}. The lattice QCD simulations are typically performed with a number of non-physical constraints; chief among them is the neglect of disconnected quark contributions.
\par
It is this feature which is considered herein for the first time. We compute the full neutral pion correlator, including disconnected contributions by using stochastic methods~\cite{Dong:1993pk,Foley:2005ac} in order to calculate estimates of the $x-x$ or \emph{all-to-all} quark propagator. The consideration of the full neutral pion and the connected pion allows disconnected terms to be isolated. This is in contrast to previous studies which have considered only the connected pion~\cite{Luschevskaya:2015cko,Bali:2017ian,Bignell:2019vpy,Bignell:2020dze,Ding:2020hxw}
\begin{align}
  \chi_{\piz^{u|d}} = \rb{ \qudb\,\gf\,\qud}.
  \label{eqn:isoSingOp}
\end{align}
\par
The magnetic polarisability can be calculated using lattice QCD by inducing a uniform external magnetic field using the background field method. This external field changes the energy of the zero-momentum neutral pion according to the relativistic energy-field relation~\cite{Lee:2014iha, QFTZuber,Luschevskaya:2015cko,Bali:2017ian,Bignell:2019vpy,Bignell:2020dze}
\begin{align}
  E^2_{\piz}\rb{B} = \mpi^2 - 4\,\pi\,\mpi\,\beta_{\piz}\,B^2 + \order{B^3}.
  \label{eqn:E2piz}
\end{align}
The extraction of the magnetic polarisability term from this energy is nontrivial as the energy contribution of the magnetic polarisability term is small compared to the overall energy of the particle~\cite{Bignell:2019vpy,Chang:2015qxa,Burkardt:1996vb,Tiburzi:2012ks,Primer:2013pva,Bignell:2020dze}. In fact, this is a requirement if the energy-field relation of \eqnr{eqn:E2piz} is to have small $\order{B^3}$ contributions. As such specialised quark operators are used to provide enhanced coupling to the energy eigenstates of the neutral pion in an external magnetic field.
\par
In this study a projection operator defined by the eigenmodes of the two-dimensional $SU(3) \times U(1)$ lattice Laplacian operator~\cite{Bignell:2020xkf} is used to project the quark sink. This enables the pion ground state in an external magnetic field to be isolated, hence allowing for an accurate determination of the magnetic polarisability of the pion. The background-field-corrected (BFC) form~\cite{Bignell:2019vpy} of the clover-fermion action~\cite{Wilson:1974sk,Sheikholeslami:1985ij} is used in order to remove the spurious background-field dependent quark mass renormalisation associated with the Wilson term~\cite{Bali:2017ian,Bignell:2019vpy}.
\par
This research is presented in the following manner. \Sec{sec:piPol} describes the chief motivation behind considering disconnected Wick contractions. A brief summary of disconnected propagator techniques is presented in \Sec{sec:props}. The details of the lattice simulations performed in this work are presented in \Sec{sec:simDet} and results in \Sec{sec:res}.
\section{Pion Polarisability}
\label{sec:piPol}
\begin{figure}
  \centering
  \begin{subfigure}[t]{\columnwidth}
    \centering
    \includegraphics[height=0.08\textheight]{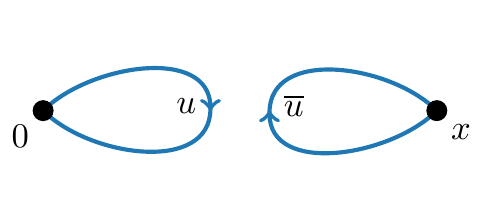}
    \includegraphics[height=0.08\textheight]{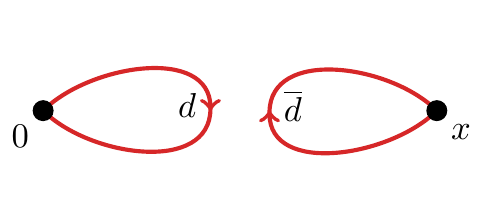}
    \includegraphics[height=0.08\textheight]{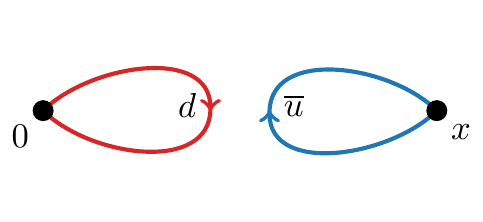}
  \end{subfigure}
  \vspace{-0.5cm}  \\
  \begin{subfigure}[t]{\columnwidth}
    \centering
    \includegraphics[height=0.08\textheight]{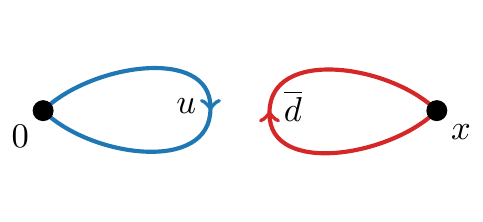}
    \includegraphics[height=0.08\textheight]{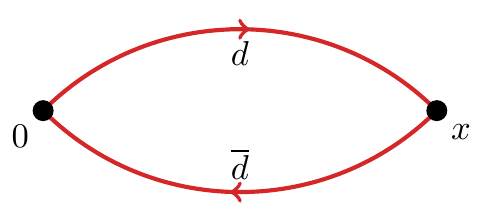}
    \includegraphics[height=0.08\textheight]{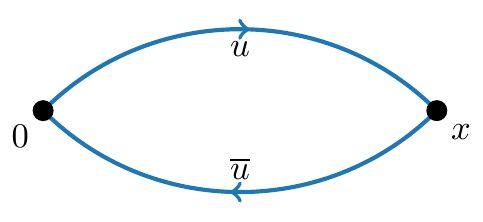}
  \end{subfigure}
  \caption{\label{fig:MesonDiscConn}Full $\piz$ meson operator Wick contractions of \eqnr{eqn:Gpiz}.}
\end{figure}
The full neutral pion interpolating operator is
\begin{align}
  \chi_{\piz} = \frac{1}{\sqrt{2}}\,\rb{ \qub\,\gf\,\qu - \qdb\,\gf\,\qd},
  \label{eqn:fullOp}
\end{align}
which produces the corresponding two-point correlation function
\begin{align}
  \mathcal{G}(t, \vec{p}) &= \frac{1}{2}\,\sum_{\vec{x}}e^{-i\vec{p}\cdot\vec{x}}\bigg[\nonumber\\   
    &-\mathrm{Tr}\big[\gf U^{ee}(x,x)\big]\mathrm{Tr}\big[\gf U^{e^{\prime}e^{\prime}}(0,0)\big]\nonumber   
    -\mathrm{Tr}\big[\gf D^{ee}(x,x)\big]\mathrm{Tr}\big[\gf D^{e^{\prime}e^{\prime}}(0,0)\big]\nonumber\\   
    &+\mathrm{Tr}\big[\gf U^{ee}(x,x)\big]\mathrm{Tr}\big[\gf D^{e^{\prime}e^{\prime}}(0,0)\big]\nonumber   
    +\mathrm{Tr}\big[\gf D^{ee}(x,x)\big]\mathrm{Tr}\big[\gf U^{e^{\prime}e^{\prime}}(0,0)\big]\nonumber\\    
    &+\mathrm{Tr}\big[\gf D^{ee^{\prime}}(x,0)\gf D^{e^{\prime}e}(0,x)\big]\nonumber 
    +\mathrm{Tr}\big[\gf U^{ee^{\prime}}(x,0)\gf U^{e^{\prime}e}(0,x)\big] \bigg],
  \label{eqn:Gpiz}
\end{align}
where colour indices are made explicit by $e$,$e^{\prime}$ and $U$ $\rb{D}$ represents an up (down) quark propagator. The disconnected quark propagators $S\rb{x,x}$ of \eqnr{eqn:Gpiz} require a source at each lattice point. The point-point $S\rb{0,0}$, point-all $S\rb{x,0}$ and all-to-all $S\rb{x,x}$ propagators are represented in \Fig{fig:MesonDiscConn}. Due to their expense, the disconnected terms are typically neglected in lattice QCD calculations.
\begin{figure}
  \centering
  \includegraphics[width=0.8\columnwidth,]{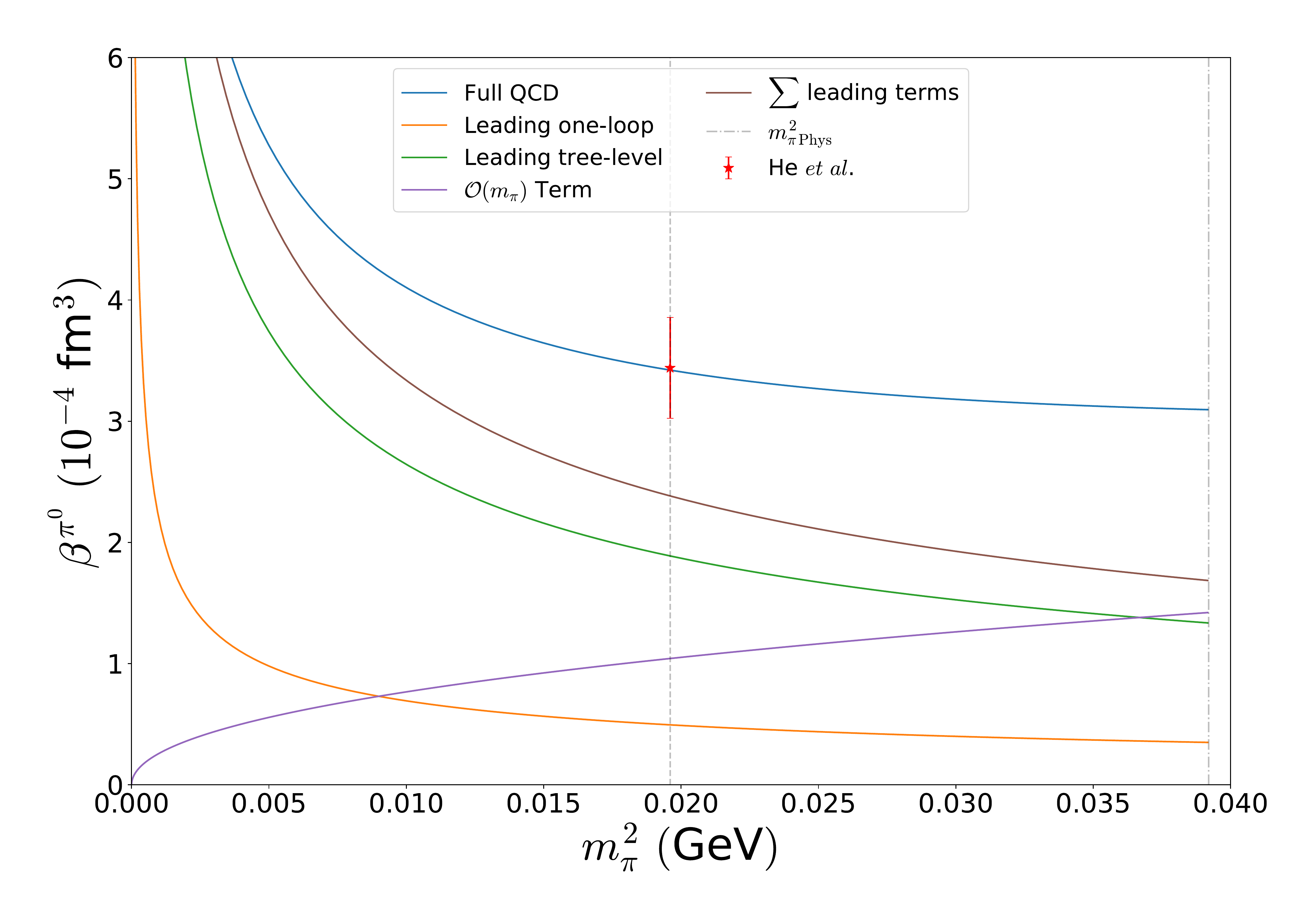}%
  \caption{\label{fig:chiEFT:loop}Leading order contributions to the magnetic polarisability of the pion as described by Eq.~20 of He \etal~\cite{He:2020ysm} for the range $0 \le m_\pi^2 \leq 2\,m_{\pi\,{\rm Phys}}^2$ where reasonable convergence is expected. The uncertainty in the result of He \etal $\,$ reflects the statistical uncertainty of the lattice results, the uncertainty in the low energy coefficients and a systematic uncertainty associated with two-loop diagrams in the electroquenched theory.}
\end{figure}
\par
In order to connect the lattice results of \Refl{Bignell:2020dze} to the physical regime, He \etal~\cite{He:2020ysm} performed a chiral extrapolation using the exact leading-order nonanalytic terms of chiral perturbation theory. Electro-quenching effects were considered through the framework of partially quenched chiral perturbation theory where it was found that the electro-quenching has no effect on the magnetic polarisability at leading loop-order. It is shown in \Refl{He:2020ysm} that the loops that contribute to the leading term are the mixed-flavour disconnected loops of \Fig{fig:MesonDiscConn}. The results of \Refl{He:2020ysm} are separated into leading loop-order and components in \Fig{fig:chiEFT:loop}. Here the leading loop-order contribution provides a $\sim 15\%$ correction at the physical point. This is a substantial proportion which must be considered in future precision-era lattice QCD studies.
\par

\begin{figure}
  \centering
  \includegraphics[width=0.8\columnwidth,]{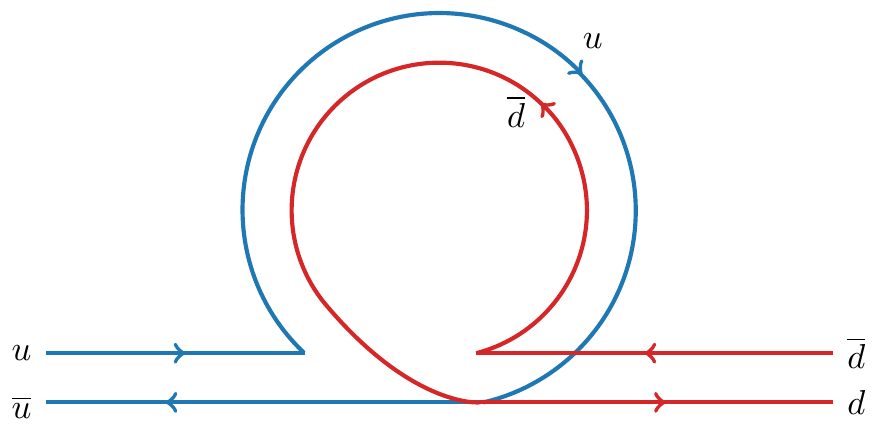}%
  \caption{\label{fig:chiEFT:uuddLoop}One-loop quark flow diagram for the neutral pion. Here we consider the Wick contraction $\mathrm{Tr}\big[\gamma_{5}D^{ee}(x,x)\big]\mathrm{Tr}\big[\gamma_{5}U^{e^{\prime}e^{\prime}}(0,0)\big]$ where the neutral pion is dressed by a $\pip$.}
\end{figure}
\par
The disconnected terms of \eqnr{eqn:Gpiz} hence pose a particularly interesting question. \Fig{fig:chiEFT:uuddLoop} illustrates how the mixed-flavour contractions can produce a $\pip$ dressing to the neutral pion. This effect is absent in the connected pion correlation functions investigated in previous studies \cite{Bali:2017ian,Bignell:2020dze,Ding:2020hxw}. This highlights the importance of including disconnected contributions to the neutral pion. In this study we will focus on contractions which produce a charged pion dressing as these are expected to be the dominant contribution at one-loop order~\cite{He:2020ysm}.
\section{Propagator Methods}
\label{sec:props}
The disconnected propagators $S\rb{x,x}$ of \eqnr{eqn:Gpiz} require a different method to conventional point-to-all propagators; here we use stochastic estimation of the matrix inverse~\cite{Dong:1993pk,Foley:2005ac}.
\par
For each correlator, we construct sets of random noise vectors $\left\{\eta\right\}$ with elements drawn from $\mbZ_4$ such that averaging over noise vectors gives
\begin{align}
  \Braket{\eta_{a\alpha}\rb{x}\,\eta^\dagger_{b\beta}\rb{y}} = \delta_{xy}\,\delta_{ab}\,\delta_{\alpha\beta},
\end{align}
where $a$, $b$ are colour indices, $\alpha$, $\beta$ are spin indices and the space-time indices are $x$, $y$. Each noise vector has a corresponding solution vector
\begin{align}
  \chi = M^{-1}\,\eta,
\end{align}
where $M$ is the fermion matrix. Hence the fermion propagator matrix element is stochastically estimated as
\begin{align}
  S_{ab;\alpha\beta}\rb{x,y} \simeq \Braket{ \chi_{a\alpha}\rb{x}\,\eta^\dagger_{b\beta}\rb{y}}.
\end{align}
The source smearing methods of \Refl{Kiratidis:2015vpa} are adopted when smearing disconnected quark propagators. The $SU(3) \times U(1)$ eigenmode projection technique of \Refl{Bignell:2020xkf} is incorporated in the $S\rb{x,x}$ propagator by applying it to the solution vector $\chi$ before the disconnected propagator calculation.
\section{Simulation details}
\label{sec:simDet}
We make use of the $2+1$ flavour dynamical QCD gauge configurations provided through the ILDG~\cite{Beckett:2009cb} by the PACS-CS~\cite{Aoki:2008sm} collaboration. The ensemble under consideration here has a pion mass $\mpi$ of $0.296$ GeV and a lattice spacing of $0.0907(13)$ fm. The lattice volume is $L^3 \times \,T = 32^3 \times\, 64$ and 150 configurations were used in this preliminary study. The Background-Field-Corrected (BFC) clover fermion action of \Refl{Bignell:2019vpy} is used to remove spurious lattice artefacts that are introduced by the Wilson term . The BFC action uses a non-perturbatively improved clover coefficient for the QCD portion of the clover term and a tree-level coefficient for the portion deriving from the background field in order to remove the additive-mass renormalisation induced by the Wilson term~\cite{Bali:2017ian,Bignell:2019vpy}.
\par
All-to-all propagators are fully diluted in time, spin and colour indices in order to reduce the statistical variance~\cite{OCais:2004xgm}. In order to improve the signal-to-noise properties of the disconnected quark propagators, we use $2^3$ spatial dilution in a manner similar to \Refltwo{Akahoshi:2019klc}{Akahoshi:2020ojo}. $\mbZ_4$ (complex $\mathbb{Z}_2$) noise is chosen due to it's superior noise reduction properties~\cite{Bulava:2008is,Foley:2010vv,Morningstar:2011ka}. The disconnected quark propagators is computed only for $t \in \left[22, 36\right]$ where we expect ground state dominance for the connected pion correlator~\cite{Bignell:2020xkf}.
\par
We implement a background magnetic field as described in \Refl{Bignell:2019vpy} using the background field method~\cite{Smit:1986fn,Burkardt:1996vb,Davoudi:2015cba}. The field strength is governed by $k_d$ as $\aqeb = \rb{2\,\pi\,k}/\rb{N_x\,N_y\,a^2}$ where the down quark has the smallest charge magnitude $q_d$. As the background magnetic field was not included at configuration generation, these configurations are \emph{electro-quenched}; the background field exists only for the valence quarks of the hadron. However, at leading loop order in the chiral expansion, there are no electroquenching errors~\cite{He:2020ysm,Hu:2007ts}.
\section{Results}
\label{sec:res}
Due to the inherent challenges in calculating the magnetic polarisability, particularly with stochastic propagators, we consider the change in the disconnected portion of the correlator as a function of field strength rather than the full magnetic polarisability. We compute correlation functions of both the connected pion and the full neutral pion as in \eqnrtwo{eqn:isoSingOp}{eqn:fullOp} respectively. Additionally we consider the single flavour operator with disconnected contributions. This enables the separation of the connected and disconnected portions of the correlation function.\begin{figure}
  \centering
  \includegraphics[width=0.51\columnwidth,]{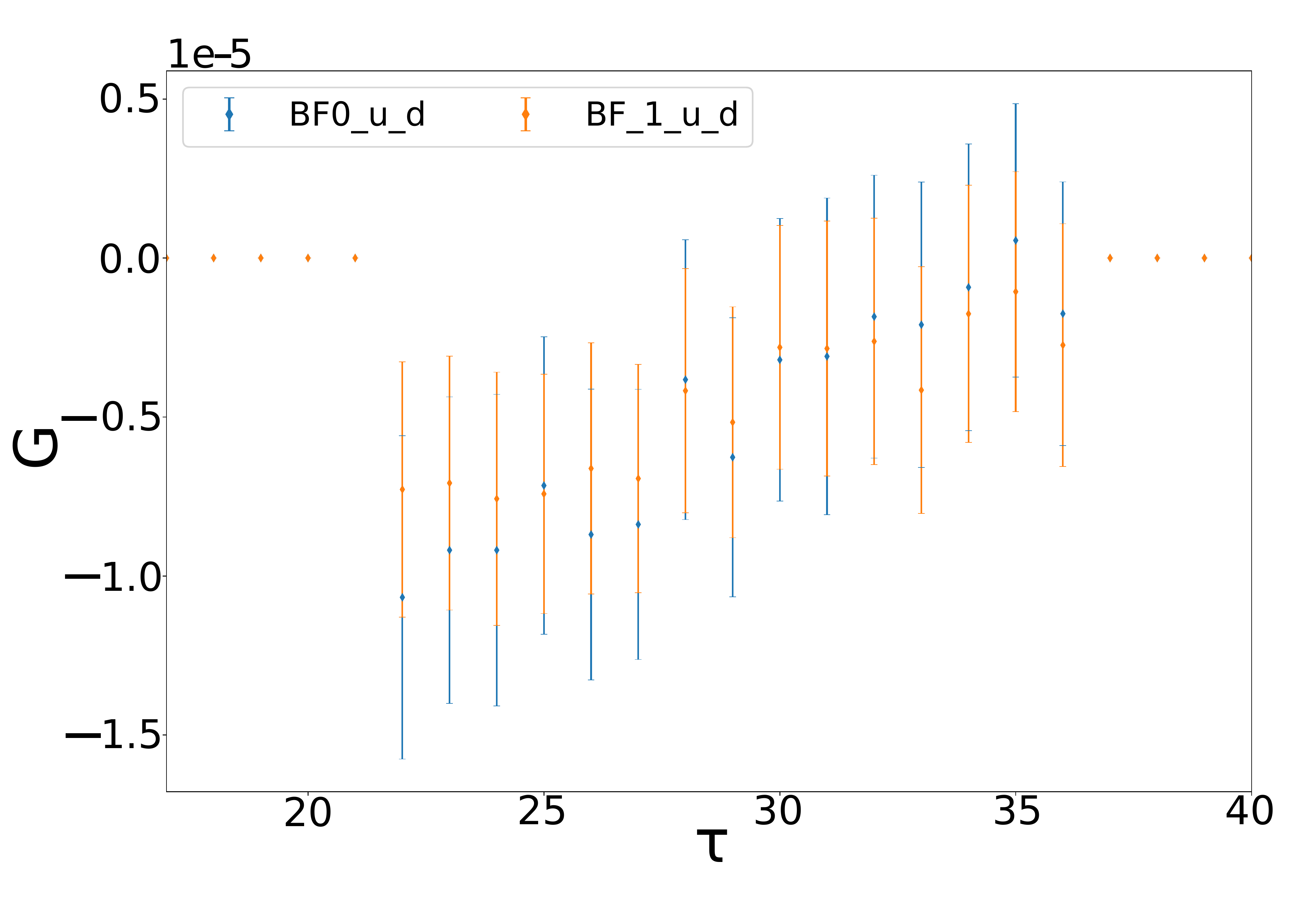}%
  \includegraphics[width=0.51\columnwidth,]{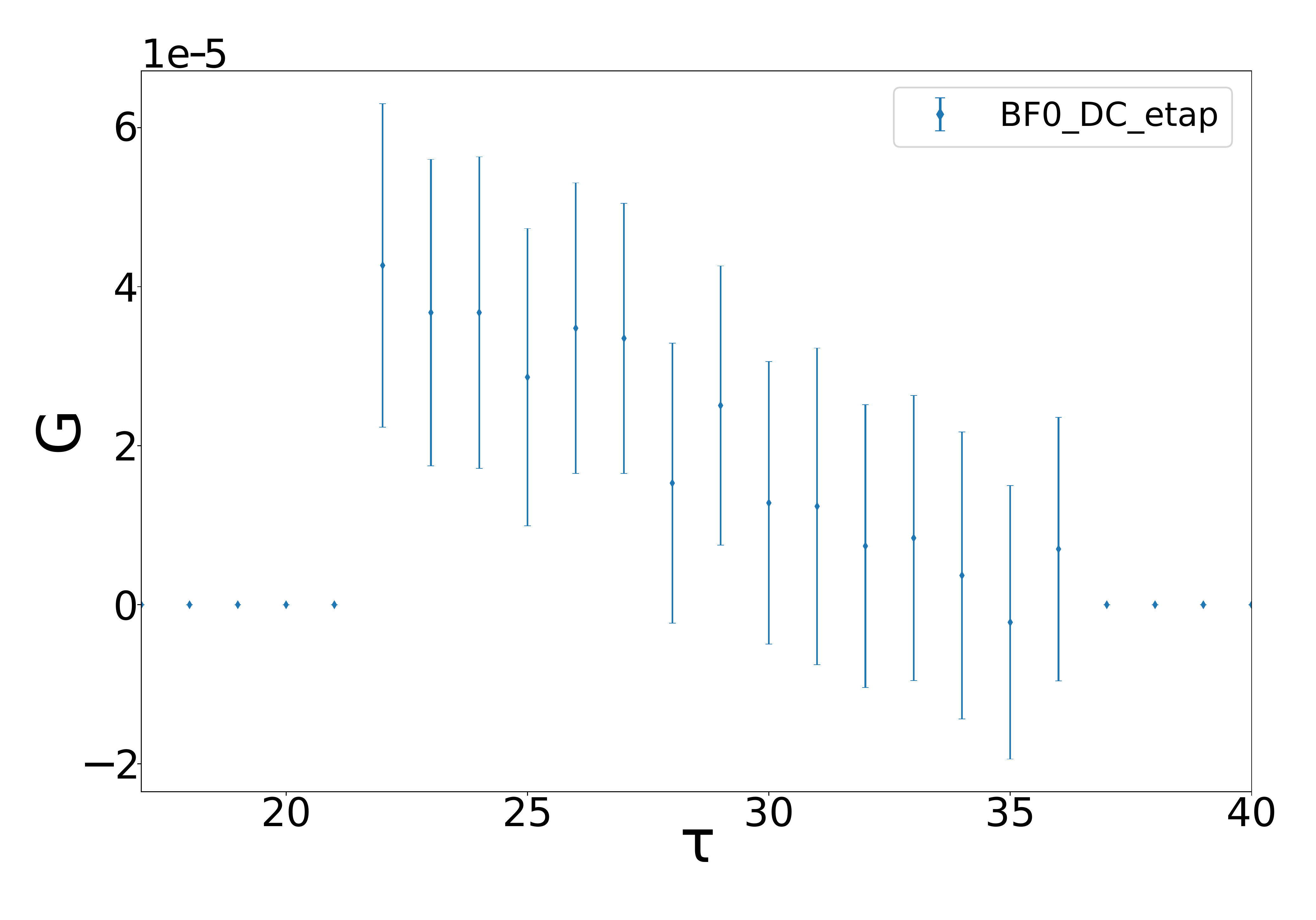}%
  \caption{\label{fig:Getap}\textbf{Left:} Disconnected $\mathrm{Tr}\big[\gf D^{ee}(x,x)\big]\mathrm{Tr}\big[\gf U^{e^{\prime}e^{\prime}}(0,0)\big]$ term with no external background field and with an external background field with $k_d = -1$. \textbf{Right:} Coherent sum of disconnected terms as appropriate for the $\eta^\prime$ for zero-background field.}
\end{figure}
\par
The disconnected mixed-flavour $u-d$ contribution to the $\piz$ correlator shown in \Fig{fig:Getap} (left) enters with a positive sign in \eqnr{eqn:Gpiz}, reflecting the negative sign between the uu and dd terms in the full $\piz$ interpolating operator. This is in contrast to the iso-scalar $\eta^\prime$
\begin{align}
    \chi_{\eta^\prime_{is}} = \frac{1}{\sqrt{2}}\,\rb{ \qub\,\gf\,\qu + \qdb\,\gf\,\qd},
\end{align}
where all the flavours of disconnected loops sum coherently and contribute with a sign opposite to the mixed flavour contributions to the pion. This acts to increase the negative slope of the correlator, hence increasing the mass of the $\eta^\prime$~\cite{Dudek:2011tt}. This is evident in \Fig{fig:Getap} (right). The iso-scalar $\eta^\prime$ is considered here in order to determine the quality of our data in a \emph{best-case} scenario where the disconnected terms sum.
\par
There is also clearly some sensitivity to the external magnetic field in \Fig{fig:Getap} (left), hence encouraging the future investigation of correlated ratios of correlators at different field strengths.

\section{Conclusion}
\label{sec:conc}
In this proceedings, we have presented the first preliminary calculation exploring the effect of a background magnetic field to the disconnected loop contractions of the neutral pion. These disconnected loops are the next-leading terms in the magnetic polarisability of the neutral pion and represent an important step towards a complete \emph{ab-initio} calculation of the magnetic polarisability of the neutral pion using lattice QCD and the background field method.
\par
The results presented in this study highlight that these disconnected contributions are in reach of today's computing efforts and represent an important opportunity to enhance our understanding of the magnetic polarisability of the neutral pion.




\begin{acknowledgments}
  R.B thanks Fangcheng He for helpful conversations. We thank the PACS-CS Collaboration for making their $2+1$ flavour configurations available via the International Lattice Data Grid (ILDG). This work was supported with supercomputing resources provided by the Phoenix HPC service at the University of Adelaide. This research was undertaken with the assistance of resources from the National Computational Infrastructure (NCI). NCI resources were provided through the National Computational Merit Allocation Scheme and supported by the Australian Government through Grant No.~LE190100021 via the University of Adelaide Partner Share. This work was supported by resources provided by the Pawsey Supercomputing Centre with funding from the Australian Government and the Government of Western Australia. This research is supported by the Australian Research Council through Grants No.~DP190102215, DP210103706, LE190100021 (D.B.L) and DP190100297 (W.K). W.K is supported by the Pawsey Supercomputing Centre through the Pawsey Centre for Extreme Scale Readiness (PaCER) program. This work is supported by STFC grants ST/T000813/1 and ST/P00055X/1 (R.B).
\end{acknowledgments}

\bibliographystyle{JHEP_arXiv}
\bibliography{DiscPion}

%

\end{document}